\documentclass[10pt,conference]{IEEEtran}

\usepackage{graphicx}
\usepackage{subfigure}
\usepackage{color}
\usepackage{epsfig}
\usepackage{amssymb}
\usepackage{amsmath}
\usepackage{amsthm}
\usepackage{latexsym}
\usepackage{setspace}
\usepackage{bbm}
\usepackage{flushend}

\newcommand{\bPP}[1]{{\mathrm{P}}_{#1}}
\newcommand{\bPr}[1]{{\mathrm{P}}\left(#1\right)}

\newcommand{\bP}[2]{\mathrm{P}_{#1}\left({#2}\right)}

\newcommand{\mc}{-\!\!\!\!\circ\!\!\!\!-}

\newcommand{\cB}{{\mathcal B}}
\newcommand{\cC}{{\mathcal C}}

\newcommand{\cK}{{\mathcal K}}

\newcommand{\cM}{{\mathcal M}}

\newcommand{\cU}{{\mathcal U}}

\newcommand{\cX}{{\mathcal X}}

\newcommand{\bF}{\mathbf{F}}

\newcommand{\bx}{\mathbf{x}}

\newcommand{\ep}{\epsilon}

\newtheorem{theorem}{Theorem}

\newtheorem*{corollary*}{Corollary}

\newtheorem{lemma}[theorem]{Lemma}
\newtheorem*{lemma*}{Lemmas}

\theoremstyle{remark}
\newtheorem*{remark*}{Remark}
\newtheorem*{remarks*}{Remarks}

\theoremstyle{definition}
\newtheorem{definition}{Definition}

\usepackage[margin=54pt]{geometry}

\begin{document}
\IEEEoverridecommandlockouts 
 
\newgeometry{top=72pt,bottom=54pt,right=54pt,left=54pt}
\title{Secret Key Capacity For Multipleaccess Channel With Public Feedback}

\author{
\IEEEauthorblockN{Himanshu Tyagi$^\ast$} 
\and
\IEEEauthorblockN{Shun Watanabe$^\dag$} 
}

\maketitle

{\renewcommand{\thefootnote}{}\footnotetext{

\vspace{.02in}\noindent

\noindent$^\ast$Department of Electrical and Computer Engineering,
 and Institute for Systems Research, University of Maryland, College
 Park, MD 20742, USA. Email: tyagi@umd.edu

\noindent$^\dag$Department of  Information Science and Intelligent Systems, 
University of Tokushima, Tokushima 770-8506, Japan, 
and Institute for Systems Research, University of Maryland, College Park,
MD 20742, USA. Email: shun-wata@is.tokushima-u.ac.jp

Himanshu Tyagi was supported by the U.S. National Science Foundation under 
Grants CCF0830697 and CCF1117546.
}}

\renewcommand{\thefootnote}{\arabic{footnote}}
\setcounter{footnote}{0}

\begin{abstract}
We consider the generation of a secret key (SK) by the inputs and the output of a secure multipleaccess
channel (MAC) that additionally have access to a noiseless public communication channel. Under specific restrictions
on the protocols, we derive various upper bounds on the rate of such SKs. Specifically, if 
the public communication consists of only the feedback from the output terminal, then 
the rate of SKs that can be generated is bounded above by the maximum symmetric rate $R_f^*$
in the capacity region of the MAC with feedback. On the other hand, if the public
communication is allowed only before and after the transmission over the MAC, then the
rate of SKs is bounded above by the maximum symmetric rate $R^*$
in the capacity region of the MAC without feedback. Furthermore, for a symmetric 
MAC, we present a scheme that generates an SK of rate $R_f^*$, improving the best previously known
achievable rate $R^*$. An application of our results establishes the SK capacity for adder MAC, 
without any restriction on the protocols. 
\end{abstract}


\section{Introduction}
What is the largest rate of a {\it secret key} (SK) that can be
generated by the inputs and the output of a secure {\it multipleaccess channel}
(MAC) with a public feedback from the output? We show that this rate
is  
bounded above by 
\begin{align}
R_f^* = \max\left\{ R : (R, R) \in \cC_{\mathtt{MACFB}} \right\},
\label{e.Rf*}
\end{align}
where $\cC_{\mathtt{MACFB}}$ denotes the capacity region\footnote{Throughout this paper, 
the capacity region of the MAC is for the average probability of error criterion.} 
of the MAC with feedback.
In fact, for a MAC that is symmetric with respect to its inputs, this largest SK
rate is equal to $R_f^*$. 

Previously, Csisz{\'a}r and Narayan \cite{CsiNar13} presented two different protocols to establish
SKs of rate 
\begin{align}
R^* = \max\left\{ R : (R, R) \in \cC_{\mathtt{MAC}} \right\},
\label{e.R*}
\end{align}
where $\cC_{\mathtt{MAC}}$ denotes the capacity region of the MAC without feedback.
In both the protocols, the inputs of the MAC were selected without any knowledge of 
the previous outputs.
Such protocols are reminiscent of SK generation in source models \cite{CsiNar04}
and will be collectively referred to as source emulation\footnote{Our source emulation
protocols include the generalized source emulation of \cite{CsiNar13, Cha10}
as a special case; the latter restricts the MAC inputs for different channel uses
to be {\it independent and identically distributed} (i.i.d.).}.
We show that $R^*$
is the best rate of an SK that can be generated using such simple protocols. 
Since for symmetric MACs 
we generate an SK of rate $R_f^*$, it follows that complex protocols that select inputs 
of the MAC based on the feedback from the output can 
outperform  source emulation. This
answers a question raised in \cite[Section VII]{CsiNar13}.

In general, the inputs of the MAC can be selected based on interactive public communication
from all the terminals after each transmission over the secure MAC. For this set-up, Csisz\'ar and
Narayan established an upper bound for the largest rate of an SK \cite{CsiNar13}, termed the SK capacity and denoted
by $C$. Moreover, for the special case of MACs in Willems class \cite{Wil82}, this upper bound was improved
and it was shown that $C \leq R_f^*$. Therefore, for symmetric MACs in Willems class, our aforementioned results
imply $C= R_f^*$. This class of channels includes adder MAC, which settles an open problem posed in \cite[Example 2]{CsiNar13}.

One of the rate $R^*$-achieving schemes in \cite{CsiNar13} involves 
transmitting messages $M_1, M_2$ of rates $(R^*,R^*)$ 
over the MAC and communicating the modulo sum $M_1\oplus M_2$
over the public channel, resulting in an SK of rate $R^*$; either $M_1$
or $M_2$ constitutes the SK. It was remarked in \cite[page 21]{CsiNar13}
that an SK generation protocol with ``{\it full feedback is ruled out 
as the feedback communication is public. Still,
if a coding scheme with partial feedback could be found by
which the gain in transmission rates exceeds the information leakage 
due to feedback, it would lead to an SK rate greater than}" $R^*$. 
Following this clue, our achievability scheme for symmetric MACs entails 
communicating 
compressed output sequences over the public channel and then extracting 
an SK of rate $R_f^*$ from the output sequence. One difficulty is the lack of 
a single-letter expression for $R_f^*$. However, this is circumvented
by converting the transmission schemes for MAC directly into SK generation protocols, without
recourse to the single-letter rate achieved. In fact, our approach implies that any
message transmission scheme of rates $(R, R)$ for a symmetric MAC 
can be used to generate an SK of rate $R$, with appropriate modifications.

Our converse proofs rely on a general converse\footnote{This general converse
is due to Prakash Narayan, who agreed to publish it in this paper.} 
for the SK generation problem in a multiterminal source model, which in turn is a simple consequence of a basic
property of interactive 
\restoregeometry
\noindent communication that was established in \cite[Lemma B.1]{CsiNar08}
(see, also, \cite{MadTet10}). Here, too, the
challenge posed by the lack of single-letter expressions is handled 
by working directly with
$n$-letter expressions.

The problem formulation and our main results are stated formally in the following section. 
Sections \ref{s.general_converse} and
\ref{s.MAC-symmetric-rate} contain the necessary tools that are
used in our converse proofs in Section \ref{s.u_bounds}.
The final section contains a discussion of our results and the properties 
of interactive
communication that are used to derive them.

\section{Problem Formulation and Main Results}\label{s.formulation_results}
Consider a MAC with two inputs\footnote{ Our results in
this paper can be extended to the multiple input case. See Section \ref{s.discussion}.}
$\cX_1$ and $\cX_2$, and an output $\cX_3$, specified by a DMC
$W: \cX_1 \times \cX_2 \rightarrow \cX_3$. We study a secrecy generation problem
for three terminals: terminals $1$ and $2$ govern the inputs 
to the DMC over which they transmit, respectively, sequences $\bx_1$ and $\bx_2$ of length $n$,
while terminal $3$ observes the corresponding $n$ length output $\bx_3$. Between
two consecutive transmissions, the terminals communicate with each other interactively
over a noiseless public communication channel of unlimited capacity. While the transmissions
over the DMC $W$ are secure, the public communication is observed by all the terminals as well as a 
(passive) eavesdropper. This model is a special case of a general model for secrecy generation 
over channels introduced by Csisz\'ar and Narayan in \cite{CsiNar13} (see also \cite{CsiNar08}).
In the manner of \cite{CsiNar13}, the messages sent over $W$ will be referred to as transmissions
and those sent over the public channel will be referred to as communication.

Formally, assume that at the outset terminal $i$ generates rv $U_i$, $i = 1, 2, 3$, to be
used for (local) randomization; the rvs $U_1, U_2, U_3$ are mutually independent.
The {\it communication-transmission protocol} can be divided into $n+1$ time slots.
In the first $n$ time slots, the terminals communicate interactively over the public channel, 
followed by a transmission over the secure DMC. The protocol ends with a final round of interactive
public communication in slot $n+1$. Specifically, in time slot $t$, $1\leq t \leq n$,
the terminals communicate interactively using their respective local randomization
$U_1, U_2, U_3$ and observations upto time slot $t-1$; the
overall interactive communication in slot $t$ is denoted by 
\begin{align}
F_t = F_t\left(U_1, U_2, U_3, X_3^{t-1}, F^{t-1}\right)
\label{e.int_comm}
\end{align}
Subsequently,
the inputs $X_{1t} = X_{1t}(F^t, U_1)$ and $X_{2t} = X_{2t}(F^t, U_2)$
are transmitted by terminals $1$ and $2$, respectively, and $X_{3t}$ is observed by 
terminal $3$. Finally, the last round of interactive communication 
$F_{n+1} = F_{n+1}\left(U_1, U_2, U_3, X_3^n, F^n\right)$ is sent
over the public channel. For convenience, we denote
$\bF = (F_1, ..., F_{n+1}).$

After the communication-transmission protocol ends, the terminals
$1$, $2$, $3$, respectively, form estimates
$K_1, K_2, K_3$ as follows:
\begin{align}
K_i= K_i(U_1, \bF), \quad i =1,2,3.
\end{align}
An rv $K$ with range $\cK$
constitutes an $\ep$-SK if the following two conditions 
are satisfied (c.f. \cite{CsiNar04}):
\begin{align}
&\hspace*{-1.5cm}\bPr{K_1 = K_2 = K_3 = K} \geq 1 - \ep, 
\label{e.recoverability}
\\
s_{in}(K; \bF) &:= \log|\cK| - H(K \mid \bF) 
\nonumber
\\
&= D\left( \bPP{K\bF} \|\bPP{\mathtt{unif}}\times \bPP{\bF}\right)
\nonumber
\\
& \leq \ep,
\label{e.security}
\end{align}
where $ \bPP{\mathtt{unif}}$ is the uniform distribution on $\cK$. 
The first condition above represents reliable {\it recoverability}
of the SK and the second guarantees its {\it security}. While our achievability
proofs establish SKs that satisfy the ``strong secrecy" condition \eqref{e.security},
our converse results are valid for SKs satisfying the weaker
secrecy condition given below:
\begin{align}
\frac{1}{n}s_{in}(K; \bF)\leq \ep.
\label{e.secrecy-weak}
\end{align} 
\begin{definition}\label{d.SKcap}
A number $R\geq 0$ is an achievable SK rate if for every $\ep > 0$, there exist local randomization $U_1, U_2, U_3$, communication-transmission
protocol $\bF$ and $\ep$-SK $K$ with
$$\frac{1}{n}\log|\cK| \geq R,$$
for all $n$ sufficiently large. 

The supremum of all achievable SK rates is called the SK capacity, denoted by $C$. 
\end{definition}
The general problem of characterizing $C$
remains open. In \cite{CsiNar13}, general lower bounds and upper bounds for $C$
were given; we state the former next, specialized for the case of two input MAC. 
\begin{theorem}\cite{CsiNar13}
The SK capacity for a MAC is bounded below as
\begin{align}
C \geq R^*.
\label{e.C_ub_CN}
\end{align}
\end{theorem}
For the special case $W(x_3\mid x_1, x_2) = \mathbbm{1}\left(x_3 = x_1 \oplus x_2\right)$, the lower bound
above is tight and $C = R^*$ \cite[Example 1]{CsiNar13}. Also, for the case when $W$ is in Willems class
of MACs \cite{Wil82}, an upper bound for $C$ was derived in \cite{CsiNar13}. 
Willems class consists of MAC where one of the
inputs, say input $1$, is determined by the output and the other input, i.e., for some mapping $\phi: \cX_2 \times \cX_3\rightarrow \cX_1$, $W(x_3\mid x_1, x_2) = 0$ if $x_1 \neq \phi(x_2, x_3)$. The following result holds.
\begin{theorem}\cite{CsiNar13}\label{t.willems_converse}
For a MAC in Willems class,
\begin{align}
C \leq R_f^*.
\label{e.C_lb_CN}
\end{align}
\end{theorem}
In this paper, we show that the bounds (\ref{e.C_ub_CN}) and (\ref{e.C_lb_CN})
are tight under various restrictions imposed on the MAC and the communication-transmission 
protocols. We first describe the specific restrictions we place. As in Definition \ref{d.SKcap},
define the SK capacity with {\it source emulation} \cite{CsiNar08, CsiNar13, Cha10}, 
denoted by $C_{\mathtt{SE}}$, as the supremum of
all achievable SK rates with the additional restriction that 
$$F_t = \text{constant}, \quad 2\leq t \leq n,$$
i.e., the transmission input sequences for the MAC are
selected solely based on the initial interactive communication
$F_1$ and local randomization at the input terminals, without any feedback from the output.
Next, define the SK capacity with {\it no input communication}, 
denoted by $C_{\mathtt{NIC}}$, as the supremum of
all achievable SK rates with the additional restriction that following the 
first round interactive communication $F_1$, the subsequent communication 
$F_2, ..., F_n$
are only from the output terminal, i.e.,
$$F_t = F_t\left(U_3, X_3^{t-1}, F^{t-1}\right), \quad 2\leq t \leq n.$$
The following inequalities ensue:
\begin{align}
C_{\mathtt{SE}} \leq C_{\mathtt{NIC}}  \leq C.
\nonumber
\end{align}

We now state our main results. First, we show a general upper bound on $C_{\mathtt{NIC}}$.
\begin{theorem}\label{t.converseNIC}
The SK capacity with no input communication is bounded above as
\begin{align}
C_{\mathtt{NIC}} \leq R_f^*.
\nonumber
\end{align}
\end{theorem}
Next, we show that for the class of symmetric MACs, this upper bound is tight.
\begin{theorem}\label{t.symmetric}
For a symmetric MAC with $\cX_1 = \cX_2$ and 
\begin{align}
W(x_3\mid x_1, x_2) = W(x_3\mid x_2, x_1),
\nonumber
\end{align}
the SK capacity with no input communication is given by
\begin{align}
C_{\mathtt{NIC}} = R_f^*.
\nonumber
\end{align}
\end{theorem}
\noindent As a corollary, we characterize $C$ for adder MAC, for which lower and 
upper bounds were reported in \cite[Example 2]{CsiNar13}.
\begin{corollary*}
For $W(x_3|x_1, x_2) = \mathbbm{1}(x_3 = x_1 + x_2)$, the SK 
capacity is given by
\begin{align}
C = R_f^*.
\nonumber
\end{align}
\end{corollary*}
\noindent Since adder MAC is in Willems class and is symmetric, the corollary 
follows from Theorem \ref{t.willems_converse} and Theorem \ref{t.symmetric}.

Finally, the following result implies that source emulation does not suffice to
generate SKs of rate $R_f^*$ and the  
complex communication-transmission protocols above are needed necessarily 
in Theorem \ref{t.symmetric}.
\begin{theorem}\label{t.capacitySE}
The SK capacity with source emulation is given by
\begin{align}
C_{\mathtt{SE}} = R^*.
\nonumber
\end{align}
\end{theorem}
\noindent The inequality $C_{\mathtt{SE}} \geq R^*$ was shown in \cite{CsiNar13}.
We show the reverse inequality in Section \ref{s.u_bounds}.
\begin{remark*}
Theorem \ref{t.capacitySE} is a further strengthening of \cite[Proposition 5]{CsiNar13}
where this result was established for source emulation protocols that restrict the
inputs of the MAC for different channel uses to be i.i.d. We show that the inequality 
$C_{\mathtt{SE}} \leq R^*$ holds even when this restriction is dropped.
\end{remark*}


\section{A General Converse for SK Capacity of a Multiterminal Source}\label{s.general_converse}
In this section, we present a converse for an SK generation problem 
in a multiterminal source model with $m$ sources (c.f. \cite{CsiNar04})
that does not require the underlying sources to be i.i.d. This specific form of
the converse is due to Prakash Narayan and it relies on a basic property of interactive
communication in multiterminal models shown in \cite{CsiNar08}. 

Terminals $1, ..., m$ observe correlated rvs $Y_1, ..., Y_m$, respectively; for brevity
we denote by $\cM$ the set $\{1, ..., m\}$  and by $Y_A$ the rvs $\{Y_i, i \in A\}$
for $A \subseteq \cM$. The terminals communicate over a public channel, possibly
interactively in several rounds. Specifically, terminal $i$ sends communication $F_{ij}$
in the $j$th round, $1 \leq j \leq r$, where $F_{ij}$ depends on the observation $Y_i$
and the previously received communication 
$$F_{11}, ..., F_{m1}, ..., F_{1j}, ..., F_{(i-1)j}.$$
We denote the overall interactive communication by $\bF$. Consider an rv $K$
taking values in $\cK$ such that 
\begin{align}
\bPr{K = K_i(Y_i, \bF), \,i \in \cM} \geq 1 -\ep,
\label{e.recoverability_sl}
\end{align}
for $0 < \ep <1$ and some mappings $K_i$ of $(Y_i, \bF)$, i.e.,
the terminals form estimates of $K$ using their respective observations $Y_i$
and the interactive communication $\bF$
that agree with $K$ with probability
greater than $1- \ep$. We present below an upper bound on $\log|\cK|$ . The following
notations will be used: Let $\cB$ be a collection of subsets of $\cM$ given by
$$\cB = \{B : B\subsetneq \cM, B \neq \emptyset \}.$$ 
A collection $\lambda = \{\lambda_B \in [0,1]: B \in \cB\}$ constitutes a fractional partition of 
$\cM$ (c.f. \cite{CsiNar08}) if
$$\sum_{B \in \cB: i \in B} \lambda_B = 1, \quad \text{for all}\,\, i \in \cM.$$
Consider a partition $\pi = \{\pi_1, ..., \pi_k\}$ of $\cM$. Corresponding to 
this partition, we define a fractional partition $\lambda^\pi$ as follows:
\begin{align}
\lambda_B^\pi = \begin{cases}
\frac{1}{k-1}, \quad &B = \pi_i^c,\,\, 1\leq i \leq k,\\
0, \quad &\text{otherwise.}
\end{cases}
\label{e.partition}
\end{align}

First, we present a key property of interactive communication 
that underlies all the converse proofs of this paper.
\begin{lemma}[{\it Interactive Communication Property}]\cite{CsiNar08}
\label{l.interactive-communication} For an interactive communication
$\bF$, we have 
$$H(\bF) \geq \sum_{B\in \cB} \lambda_B H\left(\bF\mid Y_{B^c}\right),$$
for every fractional partition $\lambda$ of $\cM$.
\end{lemma}
The following result is, in effect, a ``single-shot" converse for the SK generation problem.
\begin{theorem}\cite{Nar13}\label{theorem:one-shot-converse-source-model}
For an rv $K$ and interactive communication $\bF$ satisfying \eqref{e.recoverability_sl}, we have
\begin{align}
\log|\cK|  
\leq H\left(Y_\cM\right) - \sum_{B\in \cB}\lambda_B  H\left(Y_B\mid Y_{B^c}\right) &+ s_{in}(K; \bF)
+ \nu,
\nonumber
\end{align}
for every fractional partition $\lambda$ of $\cM$, where $\nu = (m+2)(\ep \log|\cK| + h(\ep))$.
\end{theorem}
{\it Proof.} It follows from \cite[Lemma A.2]{CsiNar08} that 
\begin{align}
H(K \mid \bF)
&\leq H\left(Y_\cM\mid \bF\right) - \sum_{B\in \cB}\lambda_B  H\left(Y_B\mid Y_{B^c},\bF\right) + \nu,
\nonumber
\\
&= H\left(Y_\cM\right) - \sum_{B\in \cB}\lambda_B  H\left(Y_B\mid Y_{B^c}\right) 
\nonumber
\\
& \quad - \left[ H(\bF)  - \sum_{B\in \cB} \lambda_B H\left(\bF\mid Y_{B^c}\right)\right] + \nu,
\nonumber
\end{align}
which, along with Lemma \ref{l.interactive-communication} and the definition of $s_{in}(K; \bF)$ in \eqref{e.security},
completes the proof.\qed
\begin{corollary*}
For $K$ and $\bF$ as in Theorem \ref{theorem:one-shot-converse-source-model}, we get
\begin{align}
\log|\cK|  
\leq \frac{1}{k-1}D\left(\bPP{Y_\cM} \middle\|\prod_{i=1}^k \bPP{Y_{\pi_i}}\right) + s_{in}(K; \bF) + \nu,
\nonumber
\end{align}
for every partition $\pi = \{\pi_1, ..., \pi_k\}$ of $\cM$.
\end{corollary*}
\noindent The corollary follows upon choosing $\lambda = \lambda^\pi$ in Theorem \ref{theorem:one-shot-converse-source-model},
where $\lambda^\pi$ is given by \eqref{e.partition}.

\section{Maximum Symmetric Rate for MAC}\label{s.MAC-symmetric-rate}

While a single-letter expression for $R^*$ is known \cite{Ahl73, Lia72},
for $R_f^*$ such an expression is available only in special cases \cite{Wil82}.
In this section, we will present $n$-letter characterizations for $R^*$ and $R_f^*$, which will
be used in our proofs in the next section. 
\begin{lemma}\label{l.R*_n_letter}
For MAC with two inputs, 
\begin{align}
R^* = \overline{\lim_n} \sup \min \bigg\{\frac{1}{n}&I\left(X_1^n \wedge X_3^n\mid X_2^n\right),
\nonumber
\\
\frac{1}{n}&I\left(X_2^n \wedge X_3^n\mid X_1^n\right), 
\nonumber
\\
\frac{1}{2n}&I\left(X_1^n, X_2^n \wedge X_3^n \right)
\bigg\},
\nonumber
\end{align}
where the $\sup$ is over all distributions 
$\bPP{X_1^n X_2^n X_3^n} = \bPP{X_1^n}\, \bPP{X_2^n}\, W^n.$
\end{lemma}
\noindent We omit the proof, which is a simple consequence of the capacity region 
for a MAC \cite{Ahl73, Lia72}.

\begin{lemma}\label{l.Rf*_n_letter}
For MAC with two inputs, 
\begin{align}
R_f^* = \overline{\lim_n} \sup \min \bigg\{\frac{1}{n}&I\left(U_1 \wedge X_3^n, U_3\mid U_2\right),
\nonumber
\\
\frac{1}{n}&I\left(U_2 \wedge X_3^n, U_3\mid U_1\right), 
\nonumber
\\
\frac{1}{2n}&I\left(U_1, U_2 \wedge X_3^n, U_3\right)
\bigg\},
\label{e.Rf*_n_letter}
\end{align}
where the $\sup$ is over all joint distributions $U_1, U_2, U_3, X_3^n$
of the randomization at the terminals and the output of the MAC
that result from communication-transmission protocols with \emph{no input communication}
(as in the definition of $C_{\mathtt{NIC}}$). 
\end{lemma}

\noindent{\it Proof.} First, we claim that making additional independent common randomness $U_3$
available to the senders and the receiver does not improve the capacity region of a MAC. Indeed,
let $\bPP{\mathsf{err}}(u_3)$ be the error probability
of the MAC $W^n$ with feedback conditioned on $U_3 = u_3$.
Clearly, there exists at least one realization $u_3^*$ such that
\begin{eqnarray*}
\bPP{\mathsf{err}}(u_3^*) \le \mathbb{E}[\bPP{\mathsf{err}}(U_3)].
\end{eqnarray*}
Thus, using the encoders and decoders with $U_3=u_3^*$ fixed we
can achieve the same rate as that of the original scheme. In the remainder of the
proof, without loss of generality, we will assume the availability of rv $U_3$
to the senders and the receiver of the MAC.

If $(R, R) \in \cC_{\mathtt{MACFB}}$, then using standard manipulations and Fano's inequality we get
\begin{align}
R \leq \frac{1}{n}I\left(U_1 \wedge X_3^n, U_3 \mid U_2\right) + \eta_n,
\nonumber
\end{align}
where $U_1, U_2$ are the messages sent by terminal $1$ and $2$, respectively,
i.i.d. uniform over $\{1, ..., \lfloor2^nR\rfloor\}$, and $\eta_n \rightarrow 0$
as $n \rightarrow \infty$. Also,
\begin{align}
R \leq \frac{1}{n}I\left(U_2 \wedge X_3^n, U_3 \mid U_1\right) + \eta_n,
\nonumber
\end{align}
and
\begin{align}
2R \leq \frac{1}{n}I\left(U_1, U_2 \wedge X_3^n, U_3\right) + \eta_n.
\nonumber
\end{align}
Since a code for MAC with feedback constitutes a valid communication-transmission
protocol with local randomization $U_1, U_2, U_3$ at terminals $1$, $2$, $3$, respectively,
it follows that $R_f^*$ is bounded above by the right-side of (\ref{e.Rf*_n_letter}). 

For the other direction, consider a MAC 
$W^{(n)}: \cU_1\times \cU_2 \rightarrow \cX_3^n\times \cU_3$
given by  
\begin{align}
&W^{(n)}\left(x_3^n, u_3\mid u_1, u_2\right) 
\nonumber
\\
&= \bPr{X_3^n = x_3^n, U_3 = u_3\mid U_1 = u_1, U_2 = u_2}.
\nonumber
\end{align}
Then, by \cite{Ahl73} and \cite{Lia72}, the right-side of (\ref{e.Rf*_n_letter})
is less than the maximum symmetric rate of the messages that can be
transmitted reliably over this MAC (without feedback). To complete the proof we note that
we can simulate $W^{(n)}$ by using the MAC $W$ with feedback $n$ times. Specifically, 
given a communication-transmission protocol with \emph{no input communication} and 
fixed values $u_1, u_2, u_3$, choosing 
\begin{align}
X_{1t} &= X_{1t}\left(u_1, F^{t-1}\left(x_3^{t-1}, u_3\right)\right),
\nonumber
\\
X_{2t} &= X_{2t}\left(u_2, F^{t-1}\left(x_3^{t-1}, u_3\right)\right),\quad 1\leq t \leq n.
\nonumber
\end{align}
simulates 
$W^{(n)}$. This is a valid choice of inputs  since both the senders 
know the common randomness $U_3$ and the
feedback signals $X_3^{t-1}$ at time $t$. \qed

\section{Upper Bounds}\label{s.u_bounds}
In this section, we prove upper bounds on $C_{\mathtt{NIC}}$
and $C_{\mathtt{SE}}$ by applying the results developed in
Sections \ref{s.general_converse} and \ref{s.MAC-symmetric-rate}.
We assume that the SK satisfies the ``weak secrecy" condition \eqref{e.secrecy-weak}.

The following observation from \cite{TyaNar13ii} is needed.
\begin{lemma} \label{lemma:factorization-preservation}
For mutually independent rvs $Y_1, Y_2, Y_3$ and an interactive communication
$\bF$ for the sources $Y_1, Y_2, Y_3$ described in Section \ref{s.general_converse},
we have
\begin{eqnarray*}
\bP{Y_1, Y_2, Y_3\mid \bF}{y_1, y_2, y_3\mid \mathbf{f}} = \prod_{i=1}^3\bP{Y_i\mid \bF}{y_i\mid \mathbf{f}},\quad \forall\,\, \mathbf{f},
\end{eqnarray*}
i.e., independent observations remain independent when conditioned on an interactive communication.
\end{lemma}

We first remark that the initial round of interactive communication
$F_1$ does not help. Specifically, for an $\ep$-SK $K$
recoverable from an interactive communication $\bF$,
it follows from \eqref{e.recoverability} and \eqref{e.security}
that there exists a fixed value $f_1$
of $F_1$ such that
\begin{align}
\bPr{K_1 = K_2 = K_3 = K\mid F_1 = f_1} &\geq 1 - 2\ep, 
\nonumber
\\
\log|\cK| - H\left(K \mid \bF, F_1 = f_1\right) &\leq 2\ep
\label{e.security_fixed_f1}
\end{align}
Note that by Lemma \ref{lemma:factorization-preservation}
the rvs
$U_1,U_2,U_3$ are conditionally independent given $F_1$.
Consider a modified protocol obtained by fixing $F_1= f_1$ 
and using local randomization $\tilde{U}_1, \tilde{U}_2, \tilde{U}_3$ 
with the same distribution as the conditional distribution of 
$U_1,U_2,U_3$ given $F_1 = f_1$. Then, in view of 
\eqref{e.security_fixed_f1}, the modified protocol generates a $2\ep$-SK
of rate not less than the original protocol and does not require any 
initial interactive communication. Thus, without loss of generality, 
in the remainder of the section we assume that $F_1$ is constant.

\subsection{Proof of $C_{\mathtt{NIC}} \leq R_f^*$}
Let $R$ be an achievable SK rate for a MAC with \emph{no input communication}. 
Setting $Y_1 = U_1$, $Y_2 = U_2$ and $Y_3 = \left(X_3^n, U_3\right)$ and applying the corollary to 
Theorem \ref{theorem:one-shot-converse-source-model}
with  partition $\pi = (\{1\}, \{2,3\} )$, 
for every $\delta>0$ and $n$ sufficiently large we have
\begin{align}
R &\leq \frac{1}{n} 
 D\left(\bPP{U_1U_2X_3^nU_3} \| \bPP{U_1} \times \bPP{U_2X_3^nU_3}\right) +\delta
\nonumber
\\
&=  \frac{1}{n} I\left(U_1 \wedge U_2, X_3^n,U_3\right)  + \delta
\nonumber
\\
&= \frac{1}{n} I\left(U_1 \wedge  X_3^n, U_3\mid U_2\right)  + \delta,
\label{e.bound1}
\end{align}
and similarly, using the partition $\pi = (\{2\}, \{1,3\} )$,
\begin{align}
R \leq \frac{1}{n} I\left(U_2 \wedge  X_3^n, U_3\mid U_1\right)  + \delta.
\label{e.bound2}
\end{align}
Also, for the partition $\pi = (\{1\},\{2\},\{3\} )$, 
we get for $n$ large
\begin{align}
R &\leq \frac{1}{2n} 
D\left(\bPP{U_1U_2X_3^nU_3} \| \bPP{U_1} \times \bPP{U_2} \times \bPP{X_3^nU_3}\right)  +\delta
\nonumber
\\
&= \frac{1}{2n} I\left(U_1, U_2 \wedge  X_3^n, U_3\right)  + \delta,
\label{e.bound3}
\end{align}
where the equality uses the independence of $U_1$ and $U_2$.
Upon combining the bounds in \eqref{e.bound1} -- \eqref{e.bound3} and taking the limit $n\rightarrow \infty$, an application of Lemma \ref{l.Rf*_n_letter} yields
\begin{align}
R \leq R_f^*,
\nonumber
\end{align}
since $\delta >0 $ was arbitrary. This proves the claimed upper bound.
\qed

\begin{remark*}
Choosing $\pi = (\{1,2\}, \{3\} )$, 
we also get the bound
\begin{align}
R &\leq \frac{1}{n} D\left(\bPP{U_1U_2X_3^nU_3} \| \bPP{U_1U_2} \times \bPP{X_3^nU_3}\right) +\delta
\nonumber
\\
&= \frac{1}{n} I\left(U_1, U_2 \wedge  X_3^n, U_3\right)  + \delta
\nonumber
\end{align}
which is subsumed by \eqref{e.bound3}.
\end{remark*}

\subsection{Proof of $C_{\mathtt{SE}} \leq R^*$}

Let $R$ be an achievable SK rate for a MAC with \emph{source emulation}. 
Setting $Y_1 = \left(X_1^n, U_1\right)$, $Y_2 = \left(X_2^n, U_2\right)$ and
$Y_3 = \left(X_3^n, U_3\right)$, and following the steps of the previous part {\it mutatis mutandis},
we get
\begin{align}
R \leq \overline{\lim_n} \sup \min \bigg\{\frac{1}{n}&I\left(X_1^n, U_1 \wedge X_3^n, U_3\mid X_2^n, U_2\right),
\nonumber
\\
\frac{1}{n}&I\left(X_2^n, U_2 \wedge X_3^n, U_3\mid X_1^n, U_1\right), 
\nonumber
\\
\frac{1}{2n}&I\left(X_1^n, U_1, X_2^n, U_2 \wedge X_3^n, U_3\right)
\bigg\}.
\label{e.step1}
\end{align}
Note that 
\begin{align}
&I\left(X_1^n, U_1 \wedge X_3^n, U_3\mid X_2^n, U_2\right) 
\nonumber
\\
&= I\left(X_1^n, U_1 \wedge X_3^n\mid X_2^n, U_2\right) 
\nonumber
\\
&\leq I\left(X_1^n \wedge X_3^n\mid X_2^n\right),
\label{e.step2}
\end{align}
where the equality follows since $U_3$ is independent of the rest of the rvs,
and the inequality\footnote{In fact, the inequality holds with equality.} uses $U_1, U_2 \mc X_1^n, X_2^n \mc X_3^n$. 
Similarly, 
\begin{align}
I\left(X_2^n, U_2 \wedge X_3^n, U_3\mid X_1^n, U_1\right) 
\leq I\left(X_2^n \wedge X_3^n\mid X_1^n\right),
\nonumber
\end{align}
and 
\begin{align}
I\left(X_1, U_1, X_2^n, U_2 \wedge X_3^n, U_3\right) 
\leq I\left(X_1^n, X_2^n \wedge X_3^n\right),
\nonumber
\end{align}
where the rvs $X_1^n = X_1^n(U_1)$ and $X_2^n = X_2^n(U_2)$
are independent.  By Lemma \ref{l.R*_n_letter} and \eqref{e.step1}, 
the upper bound on $C_{\mathtt{SE}}$ follows. \qed 

\section{Lower Bounds}\label{s.l_bounds}

In this section, we prove Theorem \ref{t.symmetric}. Suppose $(R, R)$
lies in $\cC_{\mathtt{MACFB}}$ for a symmetric MAC. Then, there exist
encoder mappings 
\begin{align}
\tau_{1t}: \{1, ..., \lfloor2^{nR}\rfloor\}\times \cX_3^{t-1} &\rightarrow \cX_1,
\nonumber
\\
\tau_{2t}: \{1, ..., \lfloor2^{nR}\rfloor\}\times \cX_3^{t-1} &\rightarrow \cX_2,\quad 1\leq t \leq n,
\label{e.encoder}
\end{align}
and decoder mapping
\begin{align}
\rho: \cX_3^n \rightarrow \{1, ..., \lfloor2^{nR}\rfloor\}\times \{1, ..., \lfloor2^{nR}\rfloor\}
\label{e.decoder}
\end{align}
such that when messages $M_1, M_2$ are sent, where rvs $M_1$ and $M_2$ are i.i.d. uniform over $\left\{1, ..., \lfloor2^{nR}\rfloor\right\}$, the error probability satisfies
\begin{align}
\ep_n  = \bPr{\rho\left(X_3^n\right) \neq (M_1, M_2)} \rightarrow 0, 
\nonumber
\end{align}
in the limit as  $n \rightarrow \infty$. 

Using this $n$ length code, we construct a symmetric code of length $2n$
by applying
(\ref{e.encoder}) and (\ref{e.decoder}) twice as follows. Consider
rvs $\hat{M}_1$, $\hat{M}_2$, $\tilde{M}_1$, $\tilde{M}_2$ i.i.d. 
uniform over $\left\{1, ..., \lfloor2^{nR}\rfloor\right\}$.
We send inputs corresponding to messages $\hat{M}_1, \hat{M}_2$ 
in the odd time instances, and, with the roles of $\tau_{1t}$ and $\tau_{2t}$ interchanged, send 
inputs  corresponding to 
messages $\tilde{M}_1, \tilde{M}_2$ in the even time instances.
Using the outputs at the odd and even time instances to decode 
$\hat{M}_1$, $\hat{M}_2$ and $\tilde{M}_1$, $\tilde{M}_2$,
respectively, we obtain a code of rate $(R, R)$ with error probability 
bounded above by $2\ep_n$. 
Denoting by $Y_t$ the rv $(X_{3(2t-1)}, X_{3(2t)})$, $1 \leq t \leq n$, 
and letting $M_1  = (\hat{M}_1, \tilde{M}_1)$ 
and $M_2  = (\hat{M}_2, \tilde{M}_2)$, we get 
\begin{align}
&H\left(Y_{t}\mid M_1, Y^{t-1}\right) 
\nonumber
\\
&= H\left(X_{3(2t-1)}\mid \hat{M}_1, X_{31}, ...,X_{3(2t-3)}\right)
\nonumber
\\ 
&\hspace*{2em}+ H\left(X_{3(2t)}\mid \tilde{M}_1, X_{32}, ..., X_{3(2t-2)}\right)
\nonumber
\\
&= H\left(X_{3(2t)}\mid \tilde{M}_2, X_{32}, ..., X_{3(2t-2)}\right) 
\nonumber
\\
&\hspace*{2em}+ H\left(X_{3(2t-1)}\mid \hat{M}_2, X_{31}, ...,X_{3(2t-3)}\right)
\nonumber
\\
&= H\left(Y_t\mid M_2, Y^{t-1}\right),
\label{e.Hsymmetry}
\end{align}
where the second equality follows from the symmetry of the MAC. 

Next, we replace the feedback $Y_t$ with its compressed version given
the observations of the input terminals. To do this, we consider a multiple-blocks 
extension of the symmetric code above and take recourse to the result
of Slepian and Wolf \cite{SleWol73}. Specifically, let $M_{1i}, M_{2i}, Y_i^n$, $i = 1, ..., N$, be 
$N$ i.i.d. repetitions of rvs $M_1, M_2, Y^n$ above. By Slepian-Wolf theorem \cite{SleWol73},
there exist mappings 
$$F_t = F_t(Y_{t1}, Y_{t2}, ..., Y_{tN}), \quad 1 \leq t \leq n,$$ 
of rates 
\begin{align}
\frac{1}{N}\log\|F_t\| &\leq  H\left(Y_{t}\mid M_1, Y^{t-1}\right)  + \ep_n,
\nonumber
\\
&=  H\left(Y_{t}\mid M_2, Y^{t-1}\right)  + \ep_n,
\label{e.SWrate}
\end{align}
such that an observer of $(M_{11}, ..., M_{1N}, Y_1^{t-1}, ..., Y_N^{t-1})$ or 
$(M_{21}, ..., M_{2N}, Y_1^{t-1}, ..., Y_N^{t-1})$ can recover $Y_t^N$
with probability of error less than $\ep_n/n$, for all $N$ sufficiently
large. The equality in (\ref{e.SWrate}) uses (\ref{e.Hsymmetry}).
Then, using a union bound on probability of error, 
the communication-transmission
protocol corresponding to $F_1, ..., F_{n}$ allows all the terminals to recover
$(Y_1^n, ..., Y_N^n)$ with probability of error less than $\ep_n$. 
Note that the overall communication-transmission protocol now consists
of $n$ rounds of communication from terminal $3$ and $2nN$ transmissions
over the MAC. In each time slot $t$, the output terminal observing $Y_{t1}, ..., Y_{tN}$
sends $F_t$ to the input terminals. Using this communication and their local observations
$M_1^N$ and $M_2^N$, the terminals $1$ and $2$ estimate $Y_{t1}, ..., Y_{tN}$ and use the estimates
to select the inputs $X_{1(2t+1)}^N, X_{1(2t+2)}^N$ and $X_{2(2t+1)}^N, X_{2(2t+2)}^N$, respectively.

Finally, we show that for all $n, N$ sufficiently large, there exists a function $K$
of $(Y_1^n, ..., Y_N^n)$ of rate $(1/nN)\log\|K\|$ greater than 
$R-\delta$, satisfying 
$$s_{in}(K; \bF) \leq \ep.$$
Therefore, $K$ is an $\ep$-SK for $n, N$ sufficiently large, where $0< \ep < 1$
is arbitrary. It remains to find a mapping $K$
as above. By \cite[Lemma 1]{CsiNar04}, it suffices to show that 
$$\left\| \bPP{K\bF} - \bPP{\mathtt{unif}}\times \bPP{\bF}\right\| \leq 2^{-n \tau},$$
for some $\tau >0$.
Indeed, by the ``balanced coloring lemma" \cite[Lemma B4]{CsiNar04}, for $n, N$
sufficiently large, there exists such a mapping $K$ of rate 
\begin{align}
\frac{1}{nN}\log\|K\| &\geq \frac{1}{nN}H(Y_1^n, ..., Y_N^n) - \frac{1}{nN}\log\|\bF\| - \ep_n
\nonumber
\\
&\geq  \frac{1}{n}H(Y^n) - \frac{1}{n}\sum_{t=1}^nH\left(Y_{t}\mid M_1, Y^{t-1}\right) - 2\ep_n
\nonumber
\\
&= \frac{1}{n}I(Y^n\wedge M_1) - 2\ep_n
\nonumber
\\
&\geq R - \delta,
\nonumber
\end{align}
where the second inequality is by (\ref{e.SWrate}) and the previous inequality uses 
Fano's inequality. Thus, $R$ is an achievable SK rate.

\section{Discussion}\label{s.discussion}
Our proof methodology in this paper is to use the basic properties
of SKs and interactive communication to obtain upper bounds on SK rates,
and then relate these upper bounds directly 
to the maximum rates of reliable transmission over a MAC, without reducing 
them to single-letter forms. In particular, this approach 
brings out a key property of interactive communication that is
instrumental in proving the converse, namely the inequality (see Lemma \ref{l.interactive-communication})
\begin{align}
H(\bF) \geq \sum_{B\in \cB} \lambda_B H\left(\bF \mid Y_{B^c}\right).
\label{e.interactive-communication}
\end{align}
For the case of two terminals, this inequality can be written as
\begin{align}
H(\bF) \geq H(\bF\mid Y_1) + H(\bF\mid Y_2),
\label{e.interactive-communication2}
\end{align}
which is well-known in the communication complexity literature (c.f. \cite{Bra12})
as the fact that {\it external information cost is at least as much as the information cost}.
Besides \eqref{e.interactive-communication}, the only other property of interactive communication
that we use is the fact that independent observations remain so when conditioned on 
interactive communication (see Lemma \ref{lemma:factorization-preservation}).
However, for a specific choice of $\lambda$ in \eqref{e.interactive-communication}, upon rearranging the terms we get
\begin{align}
I\left(Y_B\wedge Y_{B^c}\mid \bF\right) \leq I\left(Y_B\wedge Y_{B^c}\right), \quad \text{for all }B \subseteq \cM,
\nonumber
\end{align}
which in turn implies Lemma \ref{lemma:factorization-preservation}. Thus,
(\ref{e.interactive-communication}) is the only property of interactive communication that is used in our converse proofs.
Note that \eqref{e.interactive-communication2} is indeed a characteristic of an interactive communication
and does not hold for every function of $Y_1$ and $Y_2$. For instance, for symmetrically distributed 
unbiased bits $Y_1$ and $Y_2$, and $F= Y_1\oplus Y_2$,
$$H(F) = 1 < H(F\mid Y_1) + H(F\mid Y_2) = 2.$$

Our results in this paper extend easily to MACs with multiple inputs. 
In particular, Theorems \ref{t.converseNIC} and \ref{t.capacitySE} hold
for a multi-input MAC upon defining $R^*$ and $R_f^*$ as follows:
\begin{align*}
R^* &= \max\left\{R: (R, ..., R) \in \mathcal{C}_{\mathtt{MAC}}\right\},
\\
R^*_f &= \max\left\{R: (R, ..., R) \in \mathcal{C}_{\mathtt{MACFB}}\right\}.
\end{align*}
Also, Theorem \ref{t.symmetric} holds for a multi-input MAC $W: \cX_1\times ... \times \cX_{m-1} \rightarrow \cX_m$
that satisfies
\begin{align*}
W\left(x_m \mid x_1, ..., x_{m-1}\right) = W\left(x_m \mid x_{\sigma{(1)}}, ..., x_{\sigma{(m-1)}}\right),
\end{align*}
for every permutation $\sigma$ of $\{1, ..., m-1\}$.




\section*{Acknowledgements} The authors are indebted to Prakash Narayan
for allowing us to include Theorem \ref{theorem:one-shot-converse-source-model}
in this paper. Also, the discussion on interactive communication properties in the last 
section is based on ideas developed jointly with him.

\bibliography{IEEEabrv,references}
\bibliographystyle{IEEEtranS}

\end{document}